\documentclass[english]{article}
\usepackage[T1]{fontenc}
\usepackage[latin1]{inputenc}
\usepackage{amssymb}

\makeatletter

\usepackage{babel}
\makeatother
\begin{document}
\begin{center}\textbf{\Large A mathematical criterion for single photon
sources used in quantum cryptography}\end{center}{\Large \par}

\begin{center}{\large Anirban Pathak}%
\footnote{anirban.pathak@jiit.ac.in

anirbanpathak@yahoo.co.in%
} {\large }\end{center}{\large \par}

\begin{center}{\large Department of Physics, Jaypee Institute of Information
Technology, A-10, Sector-62, Noida, Up-201307, India}\end{center}{\large \par}

\begin{abstract}
A single photon source (SPS) is very important for quantum computation.
In particular, it is essential for secured quantum cryptography. But
there is no perfect SPS in reality. Therefore, probabilistic SPS where
probability of simultaneous emission of two, three, four and more
photon is less than the emission of a single photon are used. Since
classical photon always comes in bunch, the required single photon
source must be nonclassical. In the well-known antibunched state the
rate of simultaneous emission of two photon is less than that of single
photon. But the requirement of quantum cryptography is a multiphoton
version of the antibunched state or the higher order antibunched state.
Recently we have reported a mathematical criterion for higher order
antibunching. Here we have shown that any proposal for SPS to be used
in quantum cryptography should satisfy this criterion. We have studied
four wave mixing as a possible candidate of single photon source. 
\end{abstract}

\section{Introduction}

At present we normally use RSA cryptographic technique {[}\ref{the:rsa}{]}
for secured communication. The trick behind the success of this most
popular public key crypto system lies in the fact that a classical
computer takes huge time to factorise a large number, whose factors
are two large prime numbers. But this trick will not be valid if we
can construct a scalable quantum computer. This is because a quantum
computer can use Shor's algorithm {[}\ref{the:shor}{]} to factorise
a large number in a polynomial time. This observation have intensified
the interest on infinitely secured cryptographic techniques and Bennett
and Brassard's 1984 proposal {[}\ref{the:BB84}{]} for a infinitely
secured protocol for quantum cryptography (BB84 protocol) have received
the attention of the whole community. This protocol does not require
any quantum computer, so the problem with scalability or decoherence
is not important in this context and thus it is expected that the
quantum cryptographic technique will appear in the market much before
the appearance of a quantum computer. Recent experimental observations
{[}\ref{the:C-Z-Peng}-\ref{the:L-P-Lamoureux}{]}, do indicates this
fact. For example, we can note the recent success in free space distribution
of entangled photon pairs over a noisy ground atmosphere of 13 km
{[}\ref{the:C-Z-Peng}{]}. 

The basic problem with BB84 protocol is that it needs a source that
can generate single photon on demand. This is because if it produces
some pulses with multiple photon (say N photon) and Eve take one of
them and allow rest of the photon (N-1 photon) to reach Bob, then
even after the comparison with Alice's bits Bob will not be able to
detect the existence of Eve. So the system is no more infinitely secured.
To get the advantage of infinite security we need perfect single photon
source. But in reality there does not exist any perfect single photon
source (by perfect we mean a source which never produces two or more
photon simultaneously). The single photon sources, that are available
are probabilistic in nature. A probabilistic single photon source
produces single photon in most of the time but there is a finite probability
of producing multiphoton pulse. The less is this probability the better
is the source. Keeping this in mind we can say that in a probabilistic
single photon source the probability of emission of single photon
should be greater than a two photon pulse and that must be greater
than a three photon pulse and so on. Classically photon always comes
in group, and they are called bunched. When the opposite situation
occurs, and the probability of simultaneous detection of two photon
become less than the probability of their detection after a time interval
$t$ then they are called anibunched. This is a completely nonclassical
state {[}\ref{elements of quantum optics},\ref{nonclassical}{]}
(it does not have any classical analogue). For implementation of BB84
protocol we need a multiphoton version of antibunching (i.e. a higher
order version of well known antibunching phenomenon) and that is called
higher order antibunching.

Higher order extension of the nonclassical effects have been introduced
in recent past {[}\ref{hong1}-\ref{lee2}{]}. Among these higher
order nonclassical effects higher order squeezing is studied in detail
{[}\ref{hong1},\ref{hong2},\ref{hillery},\ref{giri}{]} but the
higher order antibunching which is required for implementation of
BB84 protocol is not yet studied rigorously. Actually, Lee introduced
an inequality as the criteria for higher order nonclassical state
in a pioneering paper {[}\ref{lee1}{]} by using the negativity of
the $P$ function {[}\ref{elements of quantum optics}{]}. A nonclassical
state satisfying Lee's criteria is called higher order antibunched
state and is theoretically predicted to be observable in two photon
coherent state {[}\ref{lee1}{]} and trio coherent state {[}\ref{ba an}{]}.
But from the earlier works of Lee and others {[}\ref{lee1},\ref{lee2},\ref{ba an}{]}
physical meaning of the criteria is not clear. Recently Pathak and
Garcia {[}\ref{garcia}{]} have given a simplified condition for higher
order antibunching. In next section we have briefly described BB84
protocol. In section 3 we have derived a simple condition that a SPS,
which will be used in quantum cryptography, has to satisfy. In section
4we have shown that the condition is satisfied by pump mode of a four
wave mixing process. The last section is dedicated for concluding
remarks.

\section{BB84 Protocol}

The protocol is very simple, Alice wants to send a message to Bob
in secured manner and they have chosen photon polarized along a particular
polarization axis (say horizontal) as '0' and photons polarized along
the axis perpendicular to it as '1'. Now if she sends a horizontally
polarized photon that will mean 0 and if she sends a vertically polarized
photon that will mean 1. The choice of horizontal and vertical axes
are not unique, in fact there are infinitely many possibilities. If
somebody named Eve wants to crack the information he has to measure
this photon in a particular basis (one out of infinity, so the probability
of axis matching is almost zero) and if that basis does not coincide
with the basis of Alice and Bob, then there will be a finite probability
of getting the bit incorrect (correct). After the measurement Eve
has to reproduce the bit according to his axis, since cloning is not
allowed. This will not match with Bob's axis. So there will be a finite
probability that Bob's measurement yield a wrong result. Now, by comparing
some bits with Alice, the existence of Eve can be traced by Bob. Thus
it is infinitely secured against the attack of Eve, provided you have
a single photon source.

\section{Mathematical condition for single photon source}

The $i^{th}$ factorial moment of usual number operator is defined
as, $N^{(i)}=N(N-1)........(N-i+1)$. At first we will try to understand
the meaning of $\left\langle N^{(i)}\right\rangle $, where $\left\langle \right\rangle $
denotes the quantum average. From the operator ordering theorems it
is easy to show that \begin{equation}
a^{\dagger i}a^{i}=N^{(i)}\label{eq:ho2}\end{equation}
and thus basically we need the physical meaning of $\left\langle a^{\dagger i}a^{i}\right\rangle $
which can be understood with the help of $n-th$ order correlation
function $G^{n}$.

The $n-th$ order correlation function for an electromagnetic field
is in general defined as \begin{equation}
G^{(n)}\left(x_{1}.....x_{n},y_{m}....y_{1}\right)=\left\langle E^{-}(x_{1})...E^{-}(x_{n})E^{+}(y_{n})...E^{+}(y_{1})\right\rangle \label{eq:ho13}\end{equation}
where $x_{j}=(\mathbf{r}_{j},t_{j})$ and $y_{j}=(\mathbf{r}_{j+m},t_{j+m})$.
In case of a quantum field the average in the right hand side of (\ref{eq:ho13})
is a quantum average. Otherwise the above definition of $n-th$ order
correlation is valid in general and in quantum optics it is used to
study the higher order coherence {[}\ref{elements of quantum optics}{]}.
Now if we look at a single point then $n-th$ order correlation function
(\ref{eq:ho13}) reduces to \begin{equation}
G^{(n)}\left(x_{1}......x_{1}\right)=\left\langle E^{-}(x_{1})...E^{-}(x_{1})E^{+}(x_{1})...E^{+}(x_{1})\right\rangle =\left\langle E^{-n}(x_{1})E^{+n}(x_{1})\right\rangle .\label{eq:ho14}\end{equation}
Here the single point means that the correlation or the coherence
is observed at a particular point in space at a particular time. This
definition of single point $n-th$ order correlation function or coherence
function can alternatively be written in a normalized form as \begin{equation}
G^{(n)}\left(x_{1}......x_{1}\right)=\left\langle a^{\dagger}a^{\dagger}...a^{\dagger}a...aa\right\rangle =\left\langle a^{\dagger n}a^{n}\right\rangle .\label{eq:ho16}\end{equation}
This single point correlation function is a measure of correlation
between $n$ photons of the same mode. Therefore, $\left\langle a^{\dagger n}a^{n}\right\rangle $
is a measure of the probability of observing $n$ photons of the same
mode at a particular point in space time coordinate. 

After realising the physical meaning of $\left\langle N^{(i)}\right\rangle $
we wold like to extend it into new enqualities and extract some physical
information out of them. Let us start from our physical requirement
that a single photon source to be used in quantum cryptography has
to satisfy: The probability of emission of single photon should be
greater than a two photon pulse and that must be greater than a three
photon pulse and so on. This condition can now be written as \begin{equation}
\left\langle N_{x}^{(l+1)}\right\rangle <\left\langle N_{x}^{(l)}\right\rangle \left\langle N_{x}\right\rangle <\left\langle N_{x}^{(l-1)}\right\rangle \left\langle N_{x}\right\rangle \left\langle N_{x}\right\rangle <.....<\left\langle N_{x}\right\rangle ^{l+1}.\label{eq:ho10}\end{equation}
Therefore, under the antibunching condition if we observe $(l+1)$
photons then the probability of getting them 'one by one' is maximum
and the probability of getting all the $(l+1)$ photons at a bunch
is minimum. This is what the idea of antibunching is. If we just reverse
the direction of inequality and look for bunching of photons then
for $l-th$ order bunching the possibility of getting all $(l+1)$
photons in a bunch will be maximum. 

The idea of antibunching was introduced just in opposite to bunching
and essentially that idea is manifested here. If we observe total
$(l+m)$ number of photons it is possible to get them in different
combinations, for example we can get all the photons in a bunch or
$l$ at a bunch and $m$ in another bunch and like wise. In the nonclassical
region of antibunching the probability of getting all the photons
separately (one by one) is always maximum. 

With the help of (\ref{eq:ho10}) we can simplify the condition for
obtaining $l-th$ order antibunching as \begin{equation}
d(l)=\left\langle N_{x}^{(l+1)}\right\rangle -\left\langle N_{x}\right\rangle ^{l+1}<0.\label{eq:ho21}\end{equation}
Here we can note that $d=0$ and $d>0$ corresponds to higher order
coherence and higher order bunching (many photon bunching) respectively.
This is the condition that has to be satisfied by any candidate of
SPS.

\section{The search for single photon source}

At present we cannot produce single photon source (SPS) in true sense.
All the available SPS are probabilistic in nature. In any candidate
for probabilistic SPS the probability of getting isolated photon must
be maximum and probability of getting a pulse of two or more photon
should decrease with the increase of photon number. Thus it has to
satisfy the criterion for higher order antibunching derived in last
section. Our task is to check whether well-known optical processes
can satisfy the criteria or not. Since these states are essentially
nonclassical, we have chosen a physical system which are already known
to produce nonclassical effect. The optical process which we will
study here as a possible candidate is four wave mixing process.

\subsection{Four wave mixing process}

Four wave mixing may happen in different ways. One way is that two
photon of frequency $\omega_{1}$ are absorbed (as pump photon) and
one photon of frequency $\omega_{2}$ and another of frequency $\omega_{3}$
are emitted. The Hamiltonian representing this particular four wave
mixing process is\begin{equation}
H=a^{\dagger}a\omega_{1}+b^{\dagger}b\omega_{2}+c^{\dagger}c\omega_{3}+g(a^{\dagger2}bc\,+\, a^{2}b^{\dagger}c^{\dagger})\label{eq:hamilotonian1}\end{equation}
where $a$ and $a^{\dagger}$are annihilation and creation operators
in pump mode which satisfy $[a,a^{\dagger}]$=1, similarly $b,\, b^{\dagger}$
and $c,\, c^{\dagger}$ are annihilation and creation operators in
stokes mode and signal mode respectively and $g$ is the coupling
constant. Substituting $A=a\, e^{i\omega_{1}t},B=b\, e^{i\omega_{2}t}$
and $C=c\, e^{i\omega_{3}t}$ we can write the Hamiltonian (\ref{eq:hamilotonian1})
as \begin{equation}
H=A^{\dagger}A\omega_{1}+B^{\dagger}B\omega_{2}+C^{\dagger}C\omega_{3}+g(A^{\dagger2}BC\,+\, A^{2}B^{\dagger}C^{\dagger}).\label{eq:hamilotonian}\end{equation}
Since we know the Hamiltonian we can use Heisenberg's equation of
motion (with $\hbar=1$) \begin{equation}
\dot{A}=\frac{\partial A}{\partial t}+i[H,A]\label{eq:heisenberg}\end{equation}
and short time approximation to find out the time evolution of the
essential operators. From equation (\ref{eq:hamilotonian}) we have
\begin{equation}
\begin{array}{lcl}
[H,A] & = & -A\omega_{1}-2gA^{\dagger}BC.\end{array}\label{eq:commutation}\end{equation}
From (\ref{eq:heisenberg}) and (\ref{eq:commutation}) we have \begin{equation}
\dot{A}=iA\omega_{1}-iA\omega_{1}-i2gA^{\dagger}BC=-2igA^{\dagger}BC.\label{eq:adot}\end{equation}
 Similarly \begin{equation}
\dot{B}=-igA^{2}C^{\dagger}\label{eq:bdot}\end{equation}
and \begin{equation}
\dot{C}=-gA^{2}B^{\dagger}\label{eq:cdot}\end{equation}
We can find the second order differential of $A$ using (\ref{eq:heisenberg}
and \ref{eq:adot}-\ref{eq:cdot}) as \begin{equation}
\ddot{A}=\frac{\partial\dot{A}}{\partial t}+i[H,\dot{A}]=4g^{2}AB^{\dagger}BC^{\dagger}C-2g^{2}A^{\dagger}A^{2}B^{\dagger}B-2g^{2}A^{\dagger}A^{2}C^{\dagger}C-2g^{2}A^{\dagger}A^{2}\label{eq:adoubledot}\end{equation}
Now by substituting (\ref{eq:adot}) and (\ref{eq:adoubledot}) in
the Taylor's series expansion \begin{equation}
f(t)=f(0)+t\left(\frac{\partial f(t)}{\partial t}\right)_{t=0}+\frac{t^{2}}{2!}\left(\frac{\partial^{2}f(t)}{\partial t^{2}}\right)_{t=0}......\label{eq:taylor}\end{equation}
 we obtain \begin{equation}
A(t)=A-2igtA^{\dagger}BC+\frac{g^{2}t^{2}}{2!}[4AB^{\dagger}BC^{\dagger}C-2A^{\dagger}A^{2}B^{\dagger}B-2A^{\dagger}A^{2}C^{\dagger}C-2A^{\dagger}A^{2}]\label{eq:a(t)}\end{equation}
The Taylor series is valid when t is small, so this solution is valid
for a short time and that is why it is called short time approximation.
The above calculation is show as an example. Similarly we can find
out time evolution of $B$ and $C$ or any other creation and annihilation
operator that appears in the Hamiltonian of matter field interaction.
This is a very strong technique since this straight forward prescription
is valid for any optical process where interaction time is short.
Now we can use this solution to check whether it satisfies the condition
(\ref{eq:ho21}) or not. 

Let us start with the study of the possibility of observing first
order antibunching. From equation (\ref{eq:a(t)}) we can derive expression
for $N(t)$ and $N^{(2)}(t)$ as 

\begin{equation}
\begin{array}{lcl}
N(t) & = & A^{\dagger}A+2igt\left(A^{2}B^{\dagger}C^{\dagger}-A^{\dagger2}BC\right)+g^{2}t^{2}\left(8A^{\dagger}AB^{\dagger}BC^{\dagger}C+4B^{\dagger}BC^{\dagger}C\right)\\
 & - & g^{2}t^{2}\left(2A^{\dagger2}A^{2}B^{\dagger}B+2A^{\dagger2}A^{2}C^{\dagger}C+2A^{\dagger2}A^{2}\right)\end{array}\label{eq:N(t)}\end{equation}
and 

\begin{equation}
\begin{array}{lcl}
N^{(2)}(t) & = & A^{\dagger2}A^{2}-4igtA^{\dagger3}ABC-2igtA^{\dagger2}BC+4igtA^{\dagger}A^{3}B^{\dagger}C^{\dagger}+2igtA^{2}B^{\dagger}C^{\dagger}\\
 & + & g^{2}t^{2}\left(24A^{\dagger2}A^{2}B^{\dagger}BC^{\dagger}C++32A^{\dagger}AB^{\dagger}BC^{\dagger}C+4B^{\dagger}BC^{\dagger}C\right)\\
 & - & g^{2}t^{2}\left(4A^{\dagger4}B^{2}C^{2}+4A^{4}B^{\dagger2}C^{\dagger2}+4A^{\dagger3}A^{3}B^{\dagger}B+4A^{\dagger3}A^{3}C^{\dagger}C+2A^{\dagger2}A^{2}B^{\dagger}B+2A^{\dagger2}A^{2}C^{\dagger}C\right)\\
 & - & g^{2}t^{2}\left(4A^{\dagger3}A^{3}+2A^{\dagger2}A^{2}\right).\end{array}\label{eq:Nsquare(t)}\end{equation}
In the present study all the expectations are taken with respect to
$|\alpha>|0>|0>$ for simplification. This assumption physically means
that initially a coherent state (say, a laser) is used as pump and
before the interaction of the pump with atom, there was no photon
in $b$ or $c$ mode. Thus the pump interacts with atom and causes
excitation followed by emission. Now from (\ref{eq:N(t)}) and (\ref{eq:Nsquare(t)})
we have \begin{equation}
\mathbf{\left\langle N\right\rangle ^{2}=|\alpha|^{4}-4g^{2}t^{2}|\alpha|^{6}}\label{eq:Nexpectationsquare}\end{equation}

\begin{equation}
\mathbf{\mathbf{\mathbf{\left\langle \mathbf{N^{(2)}(t)}\right\rangle }=\mathbf{|\alpha|^{4}+g^{2}t^{2}\left(-4|\alpha|^{6}-2|\alpha|^{4}\right)}}}\label{eq:n2}\end{equation}
where $A|\alpha>=\alpha|\alpha>$.

Now using (\ref{eq:Nexpectationsquare}) and (\ref{eq:n2}) we can
show that the four wave mixing process satisfies the criterion of
antibunching (\ref{eq:ho21}) because:

\begin{equation}
\begin{array}{lcl}
d(1) & = & \left\langle N^{(2)}(t)\right\rangle -\left\langle N\right\rangle ^{2}\\
 & = & \left[|\alpha|^{4}+g^{2}t^{2}(-4|\alpha|^{6}-2|\alpha|^{4})\right]-\left[|\alpha|^{4}-4g^{2}t^{2}|\alpha|^{6}\right]\\
 & = & -2g^{2}t^{2}|\alpha|^{4}\end{array}\label{eq:d(1)}\end{equation}
is always negative. Essentially, this is a nonclassical state but
mere satisfaction of nonclassicality or antbunching is not enough
we need a source which can satisfy the condition for higher order
antibunching. So, lets see what happens in the next higher order that
is in the second order. 

For the study of calculation of second order of antibunching, we can
calculate $A^{3}(t)$ simply by multiplication and operator ordering:

\begin{equation}
\begin{array}{lcl}
A^{3}(t) & = & A^{3}-6igtA^{\dagger}A^{2}BC-6igtABC+g^{2}t^{2}\left(6A^{3}B^{\dagger}BC^{\dagger}C\right)\\
 & - & g^{2}t^{2}\left(3A^{\dagger}A^{4}B^{\dagger}B+3A^{3}B^{\dagger}B+3A^{\dagger}A^{4}C^{\dagger}C+3A^{3}C^{\dagger}C\right)\\
 & - & g^{2}t^{2}\left(3A^{\dagger}A^{4}+3A^{3}+12A^{\dagger2}AB^{2}C^{2}+12A^{\dagger}B^{2}C^{2}\right).\end{array}\label{eq:Acube(t)}\end{equation}
Then $A^{\dagger^{3}}(t)$ can be written simply as,

\begin{equation}
\begin{array}{lcl}
A^{\dagger3}(t) & = & A^{\dagger3}+6igtA^{\dagger2}AB^{\dagger}C^{\dagger}+6igtA^{\dagger}B^{\dagger}C^{\dagger}+g^{2}t^{2}\left(6A^{\dagger3}B^{\dagger}BC^{\dagger}C\right)\\
 & - & g^{2}t^{2}\left(3A^{\dagger4}AB^{\dagger}B+3A^{\dagger3}B^{\dagger}B+3A^{\dagger4}AC^{\dagger}C+3A^{\dagger3}C^{\dagger}C\right)\\
 & - & g^{2}t^{2}\left(3A^{\dagger4}A+3A^{\dagger3}+12A^{\dagger}A^{2}B^{\dagger2}C^{\dagger2}+12AB^{\dagger2}C^{\dagger2}\right).\end{array}\label{eq:Adaggercube(t)}\end{equation}
Last two equations can be used to calculate the third factorial moment
$(N^{(3)}(t))$ of number operator $N$ as

\begin{equation}
\begin{array}{lcl}
N^{(3)}(t) & = & A^{\dagger3}A^{3}-6igtA^{\dagger4}A^{2}BC-6igtA^{\dagger3}ABC+6igtA^{\dagger2}A^{4}B^{\dagger}C^{\dagger}+6igtA^{\dagger}A^{3}B^{\dagger}C^{\dagger}\\
 & + & g^{2}t^{2}\left(48A^{\dagger3}A^{3}B^{\dagger}BC^{\dagger}C+108A^{\dagger2}A^{2}B^{\dagger}BC^{\dagger}C+36A^{\dagger}AB^{\dagger}BC^{\dagger}C\right)\\
 & - & g^{2}t^{2}\left(6A^{4\dagger}A^{4}B^{\dagger}B+6A^{\dagger3}A^{3}B^{\dagger}B+6A^{\dagger4}A^{4}C^{\dagger}C+6A^{\dagger3}A^{3}C^{\dagger}C+6A^{\dagger4}A^{4}\right)\\
 & - & g^{2}t^{2}\left(6A^{\dagger3}A^{3}+12A^{\dagger5}AB^{2}C^{2}+12A^{\dagger4}B^{2}C^{2}+12A^{\dagger}A^{5}B^{\dagger2}C^{\dagger2}+12A^{4}B^{\dagger2}C^{\dagger2}\right)\end{array}\label{eq:Ncube(t)}\end{equation}
Taking expectation value with respect to the intial state we can write

\[
\begin{array}{lcl}
\left\langle N^{(3)}(t)\right\rangle  & = & A^{\dagger3}A^{3}+g^{2}t^{2}\left(-6A^{\dagger4}A^{4}-6A^{\dagger3}A^{3}\right)\\
 & = & |\alpha|^{6}-g^{2}t^{2}\left(6|\alpha|^{8}+6|\alpha|^{6}\right).\end{array}\]
On the other hand, we can calculate $\left\langle N\right\rangle ^{3}$
as

\begin{equation}
\mathbf{\left\langle N(t)\right\rangle ^{3}=|\alpha|^{6}-6g^{2}t^{2}|\alpha|^{8}}.\label{eq:Nexpectationcube}\end{equation}
By using last two equations one can easily check that pump mode photon
of four wave mixing process satisfy the criteria of antibunching of
second order (\ref{eq:ho21}). Since,

\begin{equation}
\begin{array}{lcl}
d(2) & = & \left[|\alpha|^{6}-g^{2}t^{2}(6|\alpha|^{8}+6|\alpha|^{6})\right]-\left[|\alpha|^{6}-6g^{2}t^{2}|\alpha|^{8}\right]\\
 & = & -6g^{2}t^{2}|\alpha|^{6}\end{array}\label{eq:d(2)}\end{equation}
is always negative.

\section{Conclusion}

From (\ref{eq:d(1)}) and (\ref{eq:d(2)}) we can observe that the
degree of nonclassicality increases monotonically with $|\alpha|^{2}$
(the avearge photon number before the interaction). This monotonic
increase in nonclassicality is expected to be ceased with the introduction
of higher order terms. But comparing (\ref{eq:d(1)}) and (\ref{eq:d(2)})
we can easily conclude that for the same values of $g,\, t$ and $|\alpha|$,
second order antibunching is more nonclassical than the usual first
order antibunching. In other way we can say, since $d(2)$ is more
negative than $d(1)$, so the depth of nonclassicality is more in
second order antibunching. This coincides exactly with the expected
property of higher order antibunching discussed in {[}\ref{garcia}{]}.
Finally we would like to conclude that the four wave mixing process
may be a possible source of single photon needed for quantum cryptography.\\
~

\textbf{Acknowledgement}: Author thanks his student Mr. Prakash Gupta
for his efforts to reproduce and crosscheck some of the equations
and for his interesting questions and comments. He is also thankful
to DST, India for partial financial support.

\end{document}